\documentstyle[psfig]{l-aa}    

\newcommand{\ergs}{{\rm erg}\,{\rm s}^{-1}}
\newcommand{\ergcm}{{\rm erg}\,{\rm cm}^{-2}\,{\rm s}^{-1}}

\newcommand{\lx}{L_{\rm X}}
\newcommand{\ltap}{\mathrel{\hbox{\rlap{\lower.55ex \hbox {$\sim$}}
                   \kern-.3em \raise.4ex \hbox{$<$}}}}
\newcommand{\gtap}{\mathrel{\hbox{\rlap{\lower.55ex \hbox {$\sim$}}
                   \kern-.3em \raise.4ex \hbox{$>$}}}}
\newcommand{\nh}{n_{\rm H}}
\def\ea{et~al.\ }

\begin{document}
\thesaurus{01(13.25.1; 10.7.3 NGC~6652)}
\title{Detection of X-ray bursts in the globular cluster NGC~6652}
\author{J.J.M.~in~'t Zand\inst{1} \and F. Verbunt\inst{2} \and J. Heise\inst{1} 
 \and J.M.~Muller\inst{1,3} \and A. Bazzano\inst{4} 
 \and M. Cocchi\inst{4} \and L. Natalucci\inst{4} \and P. Ubertini\inst{4} 
 }
\offprints{J.J.M.~in~'t Zand}

\institute{   Space Research Organization Netherlands, Sorbonnelaan 2, 
              3584 CA Utrecht, the Netherlands
         \and
                Astronomical Institute,
              P.O.Box 80000, 3508 TA Utrecht, the Netherlands
         \and
                BeppoSAX Science Data Center, Nuova Telespazio,
		Via Corcolle 19, 00131 Roma, Italy
         \and
                Istituto di Astrofisica Spaziale (CNR),
		Via Fosso del Cavaliere, 00133 Roma, Italy
                        }
\date{Received, accepted }   
\maketitle


\begin{abstract}
Two type I X-ray bursts were detected from a position consistent with the 
transient X-ray source in NGC~6652 with the Wide Field Camera of the BeppoSAX 
satellite, strongly suggesting that this transient
is a neutron star. Our detection brings to ten the number of X-ray sources
in globular clusters in which a type I X-ray burst has been seen, out
of twelve known bright sources.
The statistical evidence that the fraction of low-mass X-ray binaries
which contain a black hole accretor is smaller in globular clusters than
in the galactic disk is suggestive, but as yet not compelling.
\end{abstract}

\section{Introduction}

The details of the formation mechanism of low-mass X-ray binaries,
in the galactic disk and in globular clusters, are not well
understood.
It is thought that low-mass X-ray binaries evolve directly from
binaries in the galactic disk, but originate from close
stellar encounters in globular clusters in which a neutron star
captures a low-mass star via tidal energy dissipation -- a tidal capture --
or in which a neutron star takes the place of a binary member in a
three-body encounter -- an exchange collision (see reviews
by Verbunt 1993, Hut et 
al.\ 1992). It has been suggested that close stellar encounters
may also produce low-mass X-ray binaries in the galactic disk,
in open clusters (Mardling 1996). 
\nocite{ver93}\nocite{hmg+92}\nocite{mar96}

A key observation that any formation theory must answer 
is the relative frequency of low-mass X-ray
binaries with a black hole and with a neutron star as accreting object.
In the galactic disk, an increasing number of low-mass X-ray binaries
with an accreting black hole has been found in the last years.
Interestingly, {\it all} of these are transient systems, in which the X-ray 
and optical luminosities are high only during outbursts, and rather
low in the quiescent intervals between the outbursts.
Since we do not know the distribution of the durations of the
inter-outburst intervals
(the observations obviously being biased to short intervals) the total
number of low-mass X-ray binaries with a black hole cannot be estimated
accurately; but it is quite possible that their number is of the same
order of magnitude as that of low-mass X-ray binaries with a neutron star
(for reviews see Tanaka \&\ Shibazaki 1996, or Chen, Shrader \&\ Livio 1997). 
\nocite{ts96}\nocite{csl97}

In contrast, none of the X-ray sources in globular clusters has so far been
found to contain a black hole (Hut et al.\ 1992; Verbunt et al.\ 1995). 
\nocite{vbhj95}
Twelve X-ray sources in globular clusters have shown X-ray luminosities
$\lx\gtap 10^{36}\ergs$, of which seven are permanently bright, and five
transients.
Type I X-ray bursts, thought to be caused by thermonuclear flashes on
or near a neutron star identifying the accreting object unambiguously as a
neutron star, have been detected in all permanently bright 
sources in globular clusters and in the transients in Terzan~5 and Liller~1.
The nature of the accreting object in the remaining three transients,
in NGC~6440, NGC~6652 and Terzan~6 is still unknown.
If the same statistics would apply in globular clusters as in the galactic
disk, these transients have a relatively high probability of containing a black
hole compared to that of the permanently bright sources.

In this article we discuss the first time discovery
of two type I X-ray bursts in the transient X-ray source in NGC~6652.
This X-ray source was detected with HEAO-1 (Hertz \&\ Wood 1985)
and shown to be associated with the globular cluster on the basis
of the data from the ROSAT All-Sky Survey (Predehl et al.\ 1991,
Verbunt et al.\ 1995). The X-ray spectrum between 0.1-2.5 keV can
be described with a powerlaw $f_{\nu}\propto\nu^{-1.07}$ (where
$f_{\nu}$ is the energy flux), absorbed by a column of 
$\nh=6.7\times 10^{20}{\rm cm}^{-2}$ (Johnston et al.\ 1996).
\nocite{hw85}\nocite{phv91}\nocite{vbhj95}\nocite{jvh96}
The X-ray luminosity was about the same during the ROSAT Survey
in September 1990 and during the ROSAT pointing in April 1992, at 
$\lx(0.5-2.5\,{\rm keV})=1.3\times10^{36}\ergs$ (for a distance of
14.3$\,$kpc; Djorgovski 1993). This is about ten times
more luminous than observed with HEAO-1 in August 1977 to April 1978,
which is why Verbunt et al.\ (1995) classify this source as a transient.
\nocite{djo93}

The observations and data analysis are described in Sect.\ 2, 
and a brief discussion follows in Sect.\ 3.

\section{Observations and data analysis}

\begin{figure}[t]
\psfig{figure=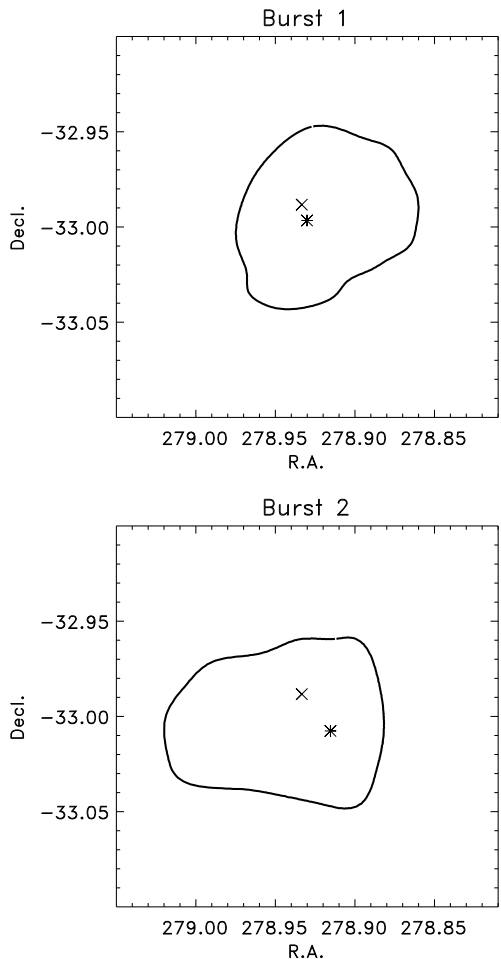,width=0.8\columnwidth,clip=t}

\caption[]{Celestial maps of 68\% confidence level regions of burst 1 and 2
for equinox 2000.0. The crosses
in both maps indicate the position of the X-ray source in NGC~6652 according
to Johnston et al. (1996) and the asterixes the best fit positions
for the BeppoSAX-WFC data \nocite{jvh96}
\label{figmaps}} 
\end{figure}

The BeppoSAX Wide Field Camera instrument (WFC, Jager et al. 1997)
consists of  two identically designed coded aperture
cameras. \nocite{jmb+97}
The main characteristics of each camera are
summarized in Table~\ref{tabwfc}. The field of view (FOV) of this
instrument is the largest of any flown X-ray imaging device. This
implies an exceptional capability to find short duration and weak
transient events.  With regards to the sensitivity it is important to
realize that an imaging device based on the coded aperture principle
has one very basic difference with direct-imaging devices such as
X-ray mirror telescopes: there is cross-talk between FOV positions
located much further from each other than the angular resolution. In
the case of WFC, there is a degradation in sensitivity to any sky
position within 20$^{\rm o}$ from a bright source. This has a
relatively large impact on observations of the crowded galactic center
field where the sensitivity is about two times less than at high
galactic latitudes far from bright sources.

\begin{table}
\caption[]{Main characteristics per BeppoSAX-WFC camera}
\begin{tabular}{ll}
\hline
Field of view           &  40$^{\rm o}$ $\times$ 40$^{\rm o}$  (3.7\% of entire sky)\\
Angular resolution      &  5 arcmin on axis\\
Source location accuracy    &  $>$0.6 arcmin (68\% conf. level)\\
Detector technology     &  Multi-wire prop. Xenon counter\\
Photon energy range     &  2 to 25 keV (read out in \\
			&  31 channels)\\
Energy resolution       &  18\% at 6~keV\\
Time resolution         &  0.5~ms\\
\hline
\end{tabular}
\label{tabwfc}
\end{table}

Detector data contain an image of the sky which is coded with the 
aperture pattern. The reconstruction of the sky image involves an
algorithm whose basic component is a cross correlation of the detector
data with the aperture pattern (Jager \ea 1997). This algorithm is 
optimum for point sources but not necessarily for diffuse sources. 
In the case of WFC the reconstruction
can be performed with an arbitrary time and photon energy 
resolution within the limitations given in Table~\ref{tabwfc}. The 
position and intensity of any point source is determined by modeling
through a point spread function (PSF). The full-width at half maximum 
of the PSF
is smallest on-axis at about 5~arcmin. The point source location accuracy
is an order of magnitude better (e.g., In~'t~Zand et al. 1997).

\begin{figure}[t]
\psfig{figure=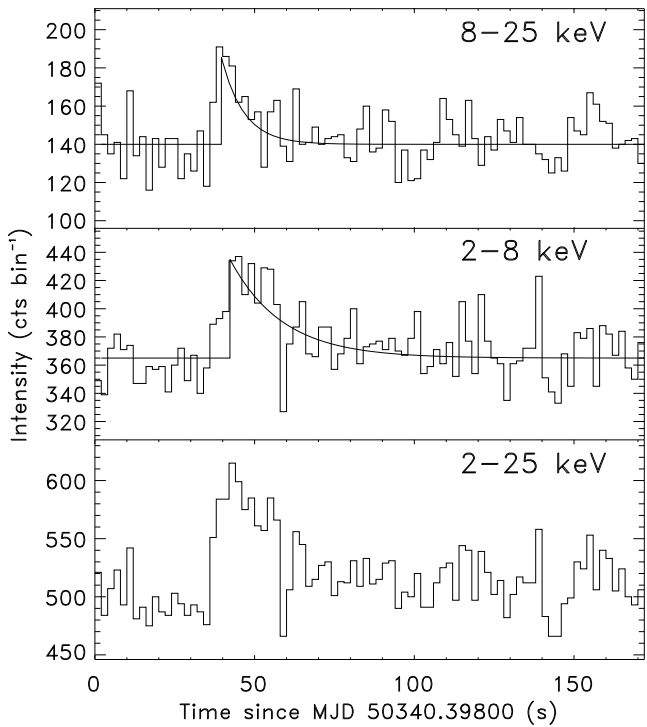,width=\columnwidth,clip=t}

\caption[]{Time profile of the first burst, per each of two bandpasses and for
the complete WFC bandpass. The bin time is 2~s. The smooth
curves are exponential models for the appropriate time profiles (see
text)
\label{figburst1}} 
\end{figure}

\begin{figure}[t]
\psfig{figure=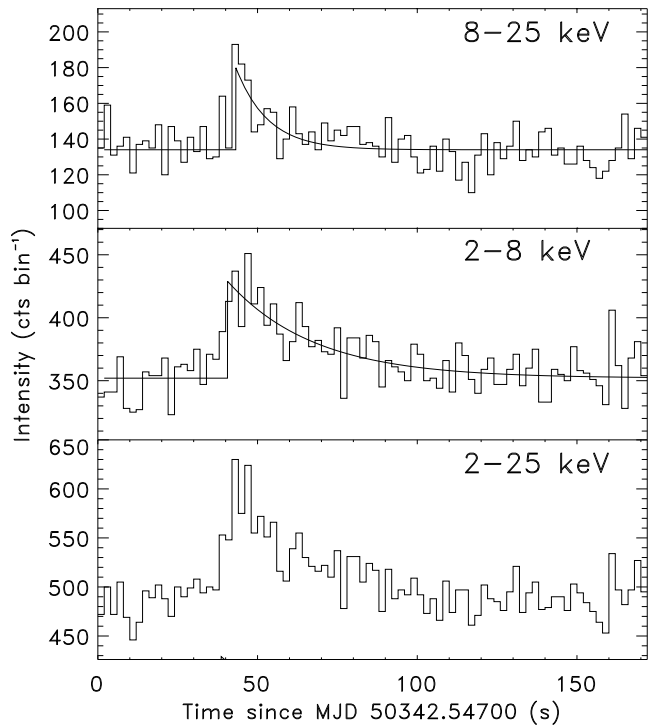,width=\columnwidth,clip=t}

\caption[]{Time profiles of the second burst
\label{figburst2}} 
\end{figure}

NGC~6652 is located 11.5~degrees from the galactic center. Therefore, it
was often observed as part of a monitoring campaign 
on the region around the galactic center. During 
the fall of 1996 
a total coverage of about $7\times10^5\,$s was obtained on this part of the sky
during about 24~days. During the spring of 1997 this was about
$4\times10^5\,$s during about 15 days. A small level of emission was 
detected from NGC~6652 in a combination of all data of 3.0$\pm$0.4~mCrab in 
2 to 8 keV (1 Crab unit is the intensity of the Crab X-ray source whose flux is
$1.8\times10^{-8}~\ergcm$ between 2 and 8 keV).

Routinely,
all data of each camera are systematically searched for burst phenomena by 
analyzing the time profile of the total detector in the full bandpass
with a time resolution of 1~s. Enhancements beyond
5$\sigma$ above a background modeled to vary linearly with time are searched 
for on time scales of up to 48~s.
For the observations discussed here this implies an on-axis sensitivity of
about 0.6~Crab units (2-25 keV) in 1~s to 0.1~Crab in 48~s.
If a burst is found,
a reconstructed sky image is generated for the time interval when the
burst occurs, another sky image is generated for a long time interval (usually
ten times as long as the burst time interval) right before the burst time, the 
latter image is normalized and subtracted from the former and this image is 
searched for point sources with intensity increases equivalent to the increase 
found in the time profile of the detector. The image subtraction is not
necessary but facilitates quick identification in case there are many
point sources in the FOV (like is the case in the observations relevant to
NGC~6652). So far, over 300 bursts were found 
and identified in all observations (Cocchi \ea 1997).
Two of these were found at a position consistent with NGC~6652. 
Fig.~\ref{figmaps} presents the error boxes of the bursts. The times,
peak intensities and best-fit positions of the two bursts are given in 
Table~\ref{tabbursts}. 

\begin{table}
\caption[]{Characteristics of two bursts}
\begin{tabular}{lll}
Parameter & Burst 1 & Burst 2 \\
\hline
Start time (MJD)	& 50340.39842 		& 50342.54744 \\
Instrument 		& WFC2	      		& WFC1		\\
Best fit R.A. (Eq. 2000.0)& 18h 35m 43s 		& 18h 35m 40s \\
Best fit Decl. 		 &-32$^{\rm o}$ 59' 48"	&-33$^{\rm o}$ 00' 27"  \\
Peak intensity (Crab, 2-8 keV) & 0.5$\pm$0.1	& 0.6$\pm$0.1 \\
Decay time $\tau$ (s, 2-8 keV)	& 16.4$\pm$5.0 		& 27.2$\pm$6.8 \\
Decay time $\tau$ (s, 8-25 keV)	& 7.0$\pm$2.4		& 9.9$\pm$3.6\\
\hline
\end{tabular}
\label{tabbursts}
\end{table}

Both bursts are relatively weak, the signal-to-noise ratio of the image being
around 6. Consequently, the extraction of meaningful time-dependent spectral
decoded information is not possible. As an alternative, we analyze time 
profiles directly of the detector for the part illuminated from the sky 
position of NGC~6652 in two
photon energy bands. By imposing less constraints on the time
profile one, on the one hand, preserves the little available statistics but,
on the other hand, ends up with a time profile that includes the
combined flux from all sources that illuminate the same part of the detector.
We regard the latter disadvantage as not important because the bursts
are a coherent and clearly recognizable signal which can only be due to
a single source of emission and the identification of the whole burst with
the source is unambiguous. Figs.~\ref{figburst1} and \ref{figburst2} 
present the time profiles for the bursts.

The time
profiles were tested for the evidence of spectral changes. They were
modeled with an exponential decay function with 4 parameters: peak intensity, 
onset time of exponential, e-folding decay time $\tau$ and background level. 
The results for $\tau$ for both energy ranges 
and the peak intensity in 2-8 keV are given in Table~\ref{tabbursts}.
In both cases the ratio of the decay time between the upper and lower energy 
band is 0.4$\pm$0.2. We regard this as good evidence 
for the presence of spectral softening in bursts from NGC~6652.

The small statistical quality of the data do not permit an analysis of the 
time-dependent decoded (background-subtracted) spectrum. This is somewhat 
different for the
average spectrum over both bursts. We fitted a number of simple spectral 
models to the spectrum. The models are described by
4 parameters and only the normalization was allowed to differ between
both bursts. All
of the models fit the data equally well with a reduced $\chi^2$ value of
0.9 for 58 independent PHA bins. The data does
not allow to single out a best fit model. Nevertheless, if we assume that
a black body model applies with absorption by cold interstellar matter
according to the model of Morrison \& 
McCammon (1983) the temperature is reasonably well constrained to 
$2.6\pm0.4$~keV. If the black body radiation is isotropic and the
distance is 14.3 kpc, we find a bolometric luminosity averaged over the
first 16~s of each burst of 2~10$^{38}$~ergs~s$^{-1}$ and a radius of the
emitting region of $6\pm2$~km.

\section{Discussion}

There are two types of X-ray bursts with durations less than a few minutes
(see review by Lewin, Van Paradijs \& Taam 1995).\nocite{lpt95}
Type II bursts have only been seen from the rapid burster and 
GRO~J1744-28 (Kouveliotou et al.\ 1996) and have spectra similar to
the underlying persistent emission. \nocite{kpf+96} They are
thought to be due to accretion instabilities.
Type I bursts are attributed to thermonuclear flashes on or near a
neutron star surface. Detection of type I bursts is, therefore, 
a strong indicator for a neutron star. One diagnostic clearly distinguishes 
type I from type II bursts: spectral softening is only seen in type I bursts.
The two bursts reported here are identified as type I bursts. 
The average black body temperature during both bursts
is consistent with such measurements in other type I bursts.
We conclude that there is strong evidence for the neutron star 
nature of this X-ray source. The
inferred size of the emitting region for the black body radiation is
consistent with a neutron star interpretation.

The ROSAT spectrum (see Sect.~1) translates into a 2 to 8 keV intensity of
2.8~mCrab. This is consistent with the
low level emission detected with BeppoSAX-WFC. This means that,
if NGC~6652 is an X-ray transient source, either since 1990 one happened 
to catch it three times during an on state or it 
is a long-duration transient. The latter is not uncommon for bright LMXB
galactic transients as exemplified by GRO~J1655-40 and KS~1731-260 
which have been in an on state for years (e.g., Chen et al.\ 1997).

In the galactic disk, about 100 low-mass X-ray binaries are known,
60 of which are permanently bright. Type I X-ray bursts have been detected
in about 20 of the permanent and 13 of the transient sources in the
galactic disk according to the tabulation by Van Paradijs (1995).
\nocite{vpa95}
The brighter permanent sources do not show type I X-ray bursts: in these the
thermonuclear fusion of helium into carbon is continuous rather than
in bursts (Lewin et al. 1995). \nocite{lpt95}
With the discovery of X-ray bursts in NGC~6652 X-ray bursts have
been detected in ten out of twelve bright X-ray sources in globular 
clusters.
Thus the fraction of sources in which an type I X-ray burst has been detected
is higher in globular clusters than in the disk; this may be
explained, partially by the larger observational interest in globular
cluster sources, and partially through the fact that none of the sources in
the globular clusters are in the high-luminosity range where type I bursts do
not occur. 

Amongst the about 40 transient low-mass X-ray binaries in the galactic
disk, some 7 have been shown to harbour compact stars with masses
deemed too high for a neutron star.
These systems are thought to harbour black holes, and indeed no
type I X-ray burst has been detected in any of them.
Various other transients are candidate black holes on the basis of
the spectral properties of their X-ray outbursts; as many as 70 \%\ of
the X-ray transients might harbour a black hole (see Table 4 in
Chen et al.\ 1997).
If the statistics in globular clusters were the same, three or
four black hole transients could be expected in this optimistic
estimation. The probability of finding zero would then be about 5\%,
a 2-$\sigma$ indication that the distribution in globular clusters
differs from that in the galactic disk.
These statistics are subject to uncertainties. For instance,
the estimate that 70\%\ of the X-ray transients in the galactic disk
harbour a black hole may be an over-estimate, since the spectral
characteristics are suggestive evidence, but no proof for a black hole.

The formation of low-mass X-ray binaries is different in 
globular clusters than in the disk. Tidal capture and exchange encounters
favour the capture of more massive stars, and thus of
black holes above neutron stars. One may thus predict a larger presence of
black holes in the X-ray binaries in globular clusters than in the galactic
disk. The formation mechanisms of low-mass X-ray binaries
in globular clusters are not understood well enough to allow reliable
estimates of this effect but it is interesting to note that it is
contrary to what observations suggest.

\begin{acknowledgements}
We thank the staff of the BeppoSAX {\em Satellite Operation Center} and
{\em Science Data Center} with the help in carrying out and processing
the WFC Galactic Center observations. The BeppoSAX satellite is a joint 
Italian and Dutch program.
\end{acknowledgements}


\end{document}